\documentclass[12pt,preprint]{aastex}

\newcommand{\zem}{\mbox{$z_{\rm em}$}}
\newcommand{\SNRE}{\mbox{1E~0102.2--7219}}

\begin{document}


\title{New X-ray quasars behind the Small Magellanic Cloud\altaffilmark{1}}

\author{A. Dobrzycki\altaffilmark{2}, K. Z. Stanek\altaffilmark{2},
L. M. Macri\altaffilmark{3,4}, and P. J. Groot\altaffilmark{5}}

\altaffiltext{1}{Based on observations collected at the Magellan Baade
6.5-m telescope.}

\altaffiltext{2}{Harvard-Smithsonian Center for Astrophysics, 60
Garden Street, Cambridge MA 02138, USA; adobrzycki@cfa.harvard.edu,
kstanek@cfa.harvard.edu.}

\altaffiltext{3}{Kitt Peak National Observatory, National Optical
Astronomy Observatory, 950 North Cherry Avenue, P.O. Box 26732,
Tucson, AZ 85726-6732, USA; lmacri@noao.edu.}

\altaffiltext{4}{Hubble Fellow}

\altaffiltext{5}{Department of Astrophysics, University of Nijmegen,
PO Box 9010, 6500 GL Nijmegen, The Netherlands; pgroot@astro.kun.nl.}

\shorttitle{X-ray quasars behind the SMC}
\shortauthors{Dobrzycki et al.}

\slugcomment{To appear in the Astronomical Journal, August 2003 issue}


\begin{abstract}

We present five X-ray quasars behind the Small Magellanic Cloud,
increasing the number of known quasars behind the SMC by
$\sim$40\%. They were identified via follow-up spectroscopy of
serendipitous sources from the {\em Chandra X-ray Observatory\/}
matched with objects from the OGLE database. All quasars lie behind
dense parts of the SMC, and could be very useful for proper motion
studies. We analyze X-ray spectral and timing properties of the
quasars. We discuss applications of those and other recently
discovered quasars behind the SMC to the studies of absorption
properties of the Cloud, its proper motion, and for establishing the
geometrical distance to the SMC.

\end{abstract}



\section{Introduction}

Recent developments in the observations of the Magellanic Clouds ---
optical monitoring campaigns (such as OGLE and MACHO), which provide
astrometry and variability information for a large number of objects,
and the observations with the {\em Chandra X-ray Observatory\/}, which
provide excellent positional accuracy in X-rays --- opened previously
unavailable windows for searches for quasars behind the Clouds. As a
result, in recent months there was a large increase in the number of
known quasars near and behind the LMC and SMC (Geha et al.\ 2003;
Dobrzycki et al.\ 2002, 2003).

In Dobrzycki et al.\ (2002) we presented an approach to searches for
such quasars based on matching of the serendipitous X-ray sources
found by {\em Chandra\/} with the optical information from the OGLE-II
database (Udalski et al.\ 1998; \.Zebru\'n et al.\ 2001). Excellent
source positions from {\em Chandra\/} match very well with the OGLE data,
which typically have very good seeing and low source confusion. The
method is perfectly suited for searches in the densest parts of the
Clouds, as it is very well shown by the quasars behind the LMC. While
$\sim$50 quasars are now known in the general direction of the LMC,
only a handful of them are behind the dense parts of the LMC bar, and
three of those are {\em Chandra\/} sources.

The downside of the method is the fact that {\em Chandra\/} performs pointed
observations only and therefore provides sparse sky coverage. For
example, at present there are only five {\em Chandra\/} targets that
coincide with the OGLE fields in the LMC.

In this paper, we applied our method to the quasar search behind the
Small Magellanic Cloud. There are two {\em Chandra\/} targets in the
SMC that coincide with the OGLE fields, and both are very
interesting. The first is NGC~346, which was observed for 100~ks. The
second is a {\em Chandra\/} calibration source, SNR \SNRE, which has
so far been observed 83 times. However, the individual observations of
the calibration source placed it at various off-axis angles on various
ACIS CCD chips, making the X-ray data analysis complicated.

Quasars behind nearby galaxies, such as the Clouds, are of great
astrophysical interest. First, they can provide a fixed reference
frame for the proper motion studies. Second, they can provide
background sources for the analysis of absorption in the foreground
galaxy. Last, but certainly not least, a hunt for X-ray quasars behind
the Magellanic Clouds has a potential for a very interesting
result. Under favorable conditions (X-ray bright and highly variable
quasar) the scattering of X-rays on the dust particles in the LMC or
SMC could lead to a direct, geometrical measurement of the distance to
the galaxy (for discussion of the concepts involved see, e.g.,
Tr{\"u}mper \& Sch{\"o}nfelder 1973; Paczy\'nski 1991; Klose 1994;
Predehl et al.\ 2000; and references therein).

\section{Candidate selection}

From the {\em Chandra\/} archive we retrieved all imaging (i.e.\ with
no grating inserted) observations which overlapped the OGLE fields.
We reduced and analyzed the X-ray data using tools available in the
CIAO 2.3 and SHERPA software
packages.\footnote{http://cxc.harvard.edu/ciao/} We applied newest
gain and aspect corrections to the event lists. We cleaned the
electronic streaks in ACIS-S4 chip using DESTREAK. We searched the
data for serendipitous point sources using both WAVDETECT and
CELLDETECT source detection tools. Searching for sources was
straightforward in the case of the observation of NGC~346. For \SNRE\
it was more complicated. We concentrated on those observations for
which the aim point and large part of the field of view overlapped the
OGLE fields. We ran detection tools on all individual observations,
but also, whenever it was practical (i.e.\ when aim points were close
enough --- typically within 2 arcmin --- so that the point spread
function size was not dramatically varying from observation to
observation) we merged the individual observations in an attempt to
improve the photon statistics, and ran detection tools on the merged
data sets. However, merging was possible in only a handful of
cases. We combined the outputs of all detect runs. As expected, this
process led to several spurious sources and duplicate detections. We
examined the resulting list of candidates and removed the obvious
artifacts and reconciled multiple detections.

{\em Chandra\/} X-ray positions are typically accurate to better than
1 arcsec.\footnote{http://cxc.harvard.edu/cal/ASPECT/celmon/; see also
Garmire et al.\ (2000).} For each X-ray source, we identified both the
closest OGLE object and the closest OGLE variable. We used a
conservative threshold of 5 arcsec for the position match. The
resulting list contained 80 candidates in both fields.

To aid quasar candidate selection, whenever the net number of source
X-ray photons allowed it, we performed a quick spectral analysis,
identifying sources with power-law intrinsic spectra and/or higher
than typical absorption, which are primary candidates for being
QSOs. However, this was only used for prioritizing the follow-up
observations and we did not exclude any objects based on this
characteristics.

\section{Observations and identifications}

\begin{figure}
\epsscale{0.82}
\plotone{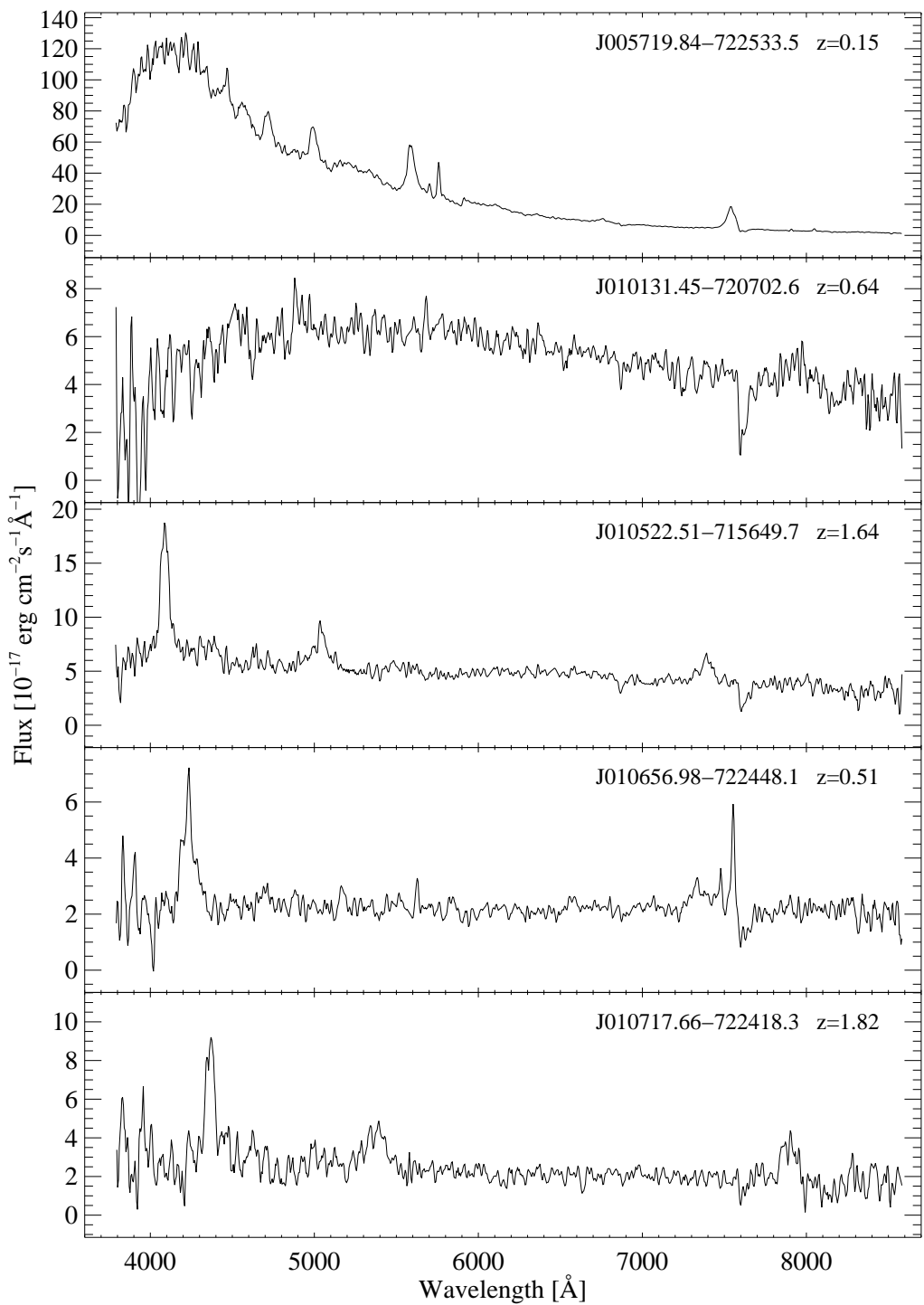}
\caption{Magellan Baade 6.5-m/LDSS-2 spectra of the identified X-ray
quasars. The spectra were smoothed with a 5-pixel
box.\label{fig:spectra}}
\end{figure}

\begin{figure}
\epsscale{1.0}
\plotone{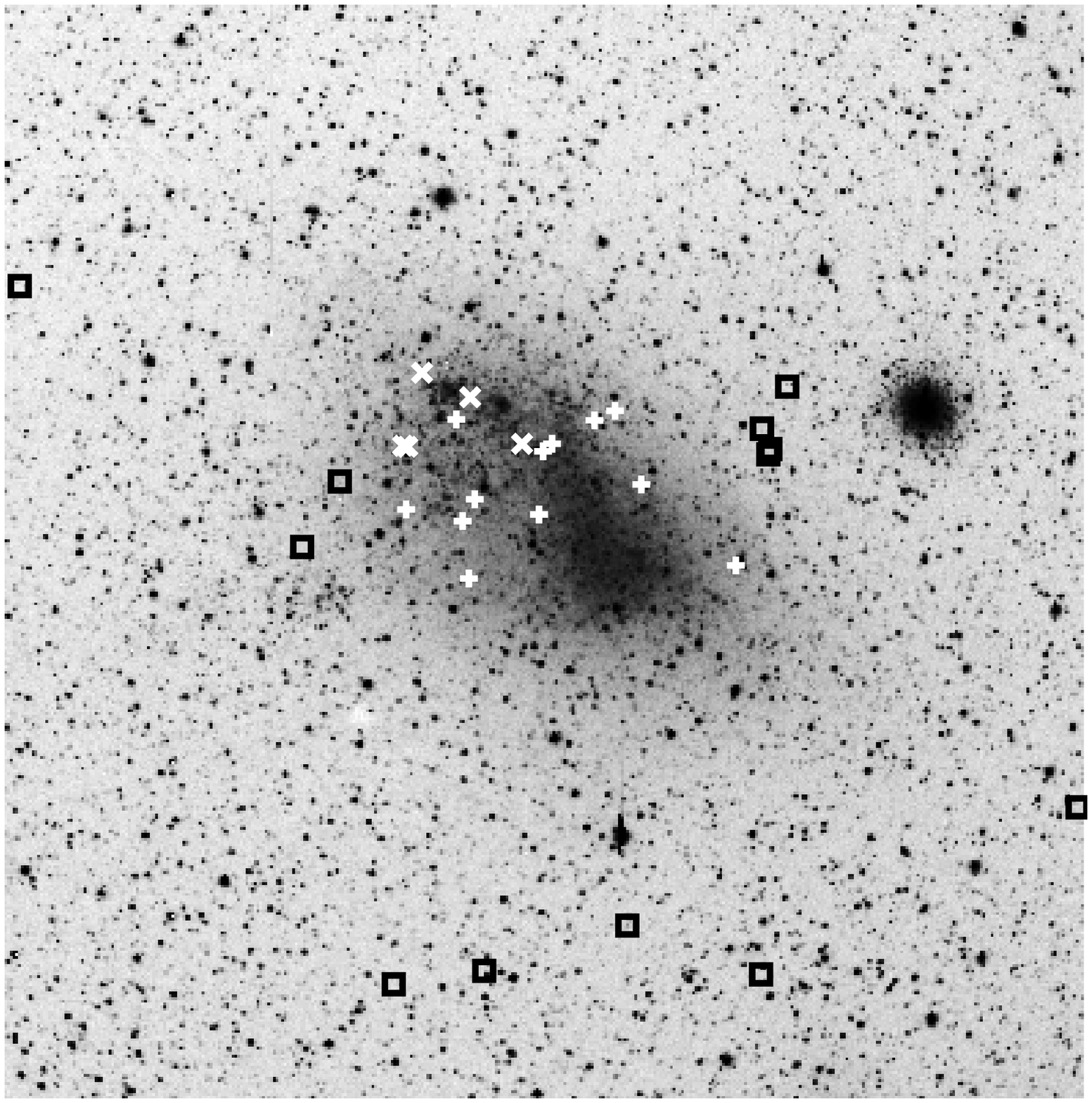}   
\caption{Quasars in the vicinity of the Small Magellanic Cloud. The
image is roughly $7^\circ\times7^\circ$. The quasars presented in
this paper are shown with white ``X'' marks. White crosses show the
variability-selected quasars from Geha et al.\ (2003) and Dobrzycki et
al.\ (2003). Black squares show the quasars known before 2002; a
dramatic improvement in recent months in both the number of known
objects and the coverage is clearly seen. SMC image courtesy of
G.~Bothun.\label{fig:smcimage}}
\end{figure}

The optical follow-up spectra were obtained on 2002 September 16-18
with the Magellan Baade 6.5-meter telescope. We used the LDSS-2
imaging spectrograph, with the 1.03~arcsec slit and the 300~l/mm
grism, yielding a nominal resolution of 13.3~\AA. Exposure times
ranged from 120 to 600 seconds. All observations were carried out with
the slit oriented in the east-west direction. Additionally, we
observed two spectrophotometric standards, LTT 1788 and LTT 7379
(Hamuy et al.\ 1992). Following each observation, a He-Ne arc lamp
spectrum was acquired for wavelength calibration purposes. Spectra
were reduced in the standard way using IRAF.

We identified five quasars among 39 observed candidates; we show their
optical spectra in Figure~\ref{fig:spectra}. We will discuss the
properties of the individual objects in detail in the following
section. We note that only one of the new quasars, QSO
J005719.84--722533.5, was identified by OGLE as a variable. This is a
clear indication that the variability- and X-ray-based quasar
selections should be considered as complementing one another.

The efficiency of the method is similar to the one seen in the LMC
(Dobrzycki et al.\ 2002). The other objects identified in the
follow-up observations were primarily early type stars and X-ray
binaries. We will present the analysis of those objects in a
forthcoming paper.

On Figure~\ref{fig:smcimage} we show the locations of the new X-ray
quasars, and the positions of the variability-selected quasars from
Geha et al.\ (2003) and Dobrzycki et al.\ (2003). We also show the
locations of quasars known prior to 2002. As it can be clearly seen,
the methods proposed in the recent papers made a significant
difference in both the coverage and the number of known objects.
Until mid-2002, there were no known QSOs behind the dense parts of the
SMC. Variability-based techniques revealed twelve such quasars. This
is still a rather low number, and with the identification of the X-ray
QSOs, the number of known quasars behind the dense parts of the SMC
has risen by more than 40\%.

\section{Object properties}

\subsection{QSO J005719.84--722533.5}

This is a relatively bright ($I=17.5$) quasar at $\zem=0.15$. It is
seen serendipitously at a large (17 arcmin) off-axis angle in a 100~ks
{\em Chandra}/ACIS observation (ObsID 1881) of NGC~346 done on 2001
May 15. As luck would have it, the quasar happened to fall on the
backside-illuminated CCD chip ACIS-S3, which has better efficiency at
low energies than the frontside-illuminated chips. It is the brightest
X-ray quasar from the objects presented here. The event pileup is not
a problem due to the large off-axis angle.

\begin{figure}[ht]
\epsscale{1.0}
\plotone{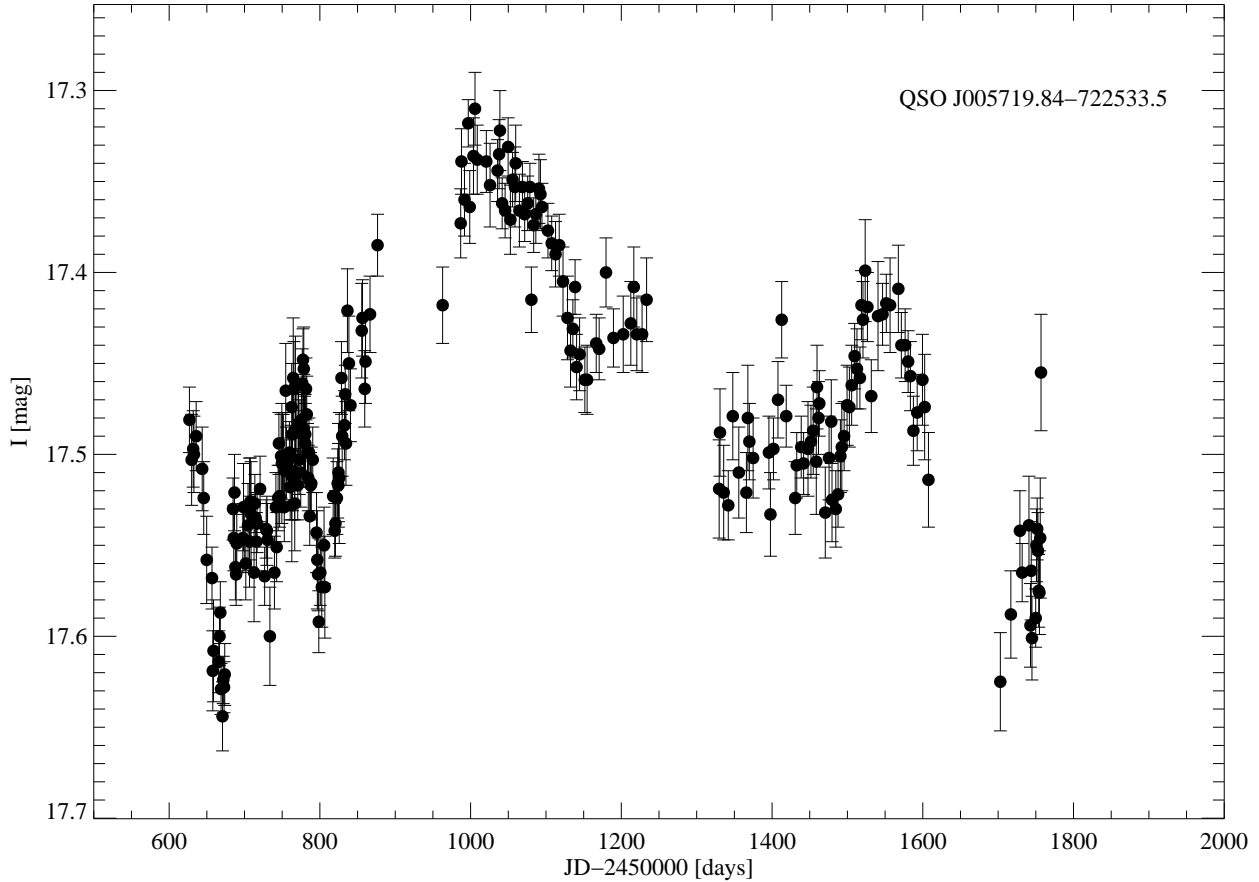}
\caption{OGLE (\.Zebru\'n et al.\ 2001) light curve of QSO
J005719.84--722533.5. JD 2,450,000 corresponds to UT 1995 October
9.\label{fig:0057ogle}}
\end{figure}

\begin{figure}
\epsscale{0.95}
\plotone{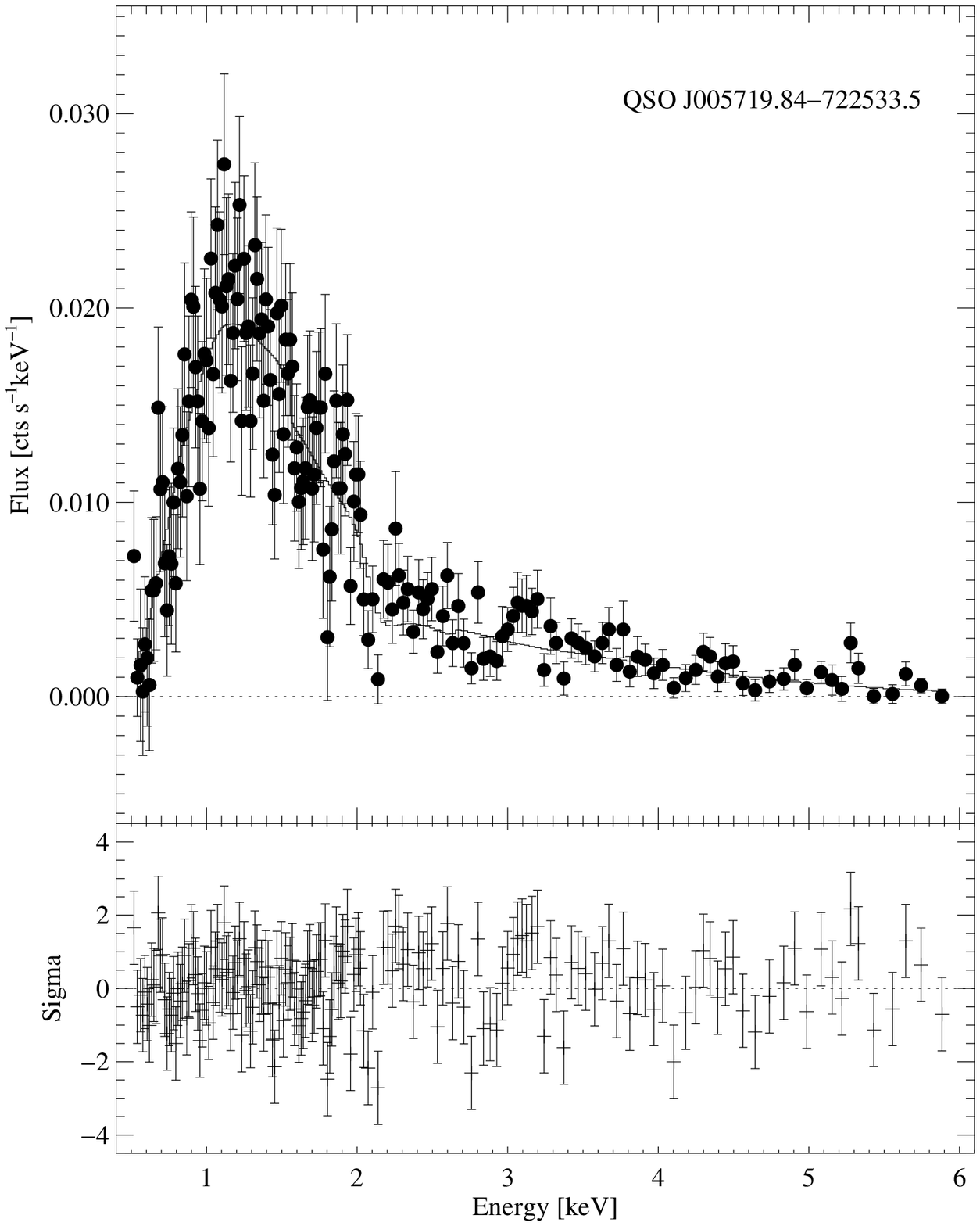}
\caption{{\em Chandra\/} ACIS-S3 spectrum of QSO J005719.84--722533.5. The
bottom panel shows the residuals to the best fit power law plus
Galactic, SMC, and quasar intrinsic absorption.\label{fig:0057xspe}}
\end{figure}

Even though the source is very far off-axis in the {\em Chandra\/}
observation, where the point spread function is very large, the X-ray
and OGLE positions agree remarkably well, to better than 1~arcsec. The
quasar lies in the vicinity of a known, but previously unclassified
X-ray source 2E~0055.6-7241 (see Table~\ref{tab:x}).

QSO J005719.84--722533.5 is the only object among the quasars
presented here which is classified in \.Zebru\'n et al.\ (2001) as
variable. Its $I$ light curve (Figure~\ref{fig:0057ogle}) clearly
shows slow irregular variability with an amplitude of
$\sim$0.2~mag. However, this object failed two initial criteria for
inclusion in the Eyer (2002) variability-based analysis: it did not
have $B$ photometry available and it was too red in the $V-I$
color. Therefore, it was not considered by Eyer to be a quasar
candidate.

The X-ray observation yielded $\sim$2800 source photons with
$E\geq0.3$ keV, sufficient to establish basic X-ray properties of the
quasar. We excluded photons with energies below 0.3~keV since there
are large calibration uncertainties for soft X-rays. We extracted
source and background spectra using DMEXTRACT.
Figure~\ref{fig:0057xspe} shows the X-ray spectrum. Significant
absorption in low energies is clearly visible.

We performed spectral fits with SHERPA. We fitted the spectrum
assuming the intrinsic quasar spectrum to be a power law and fixed
Galactic absorption towards the SMC. The Milky Way absorbing column
alone is not sufficient to explain the dip in the spectrum in the low
energies, and additional source of absorption is needed.

In Table~\ref{tab:x} we list fit results for two scenarios. First, we
attributed the additional absorption to the SMC material alone (in all
fits presented in this paper, we assumed the SMC metallicity to be
20\% solar; Russell \& Dopita 1990, 1992). In Table~\ref{tab:x} we
also list, for comparison, the SMC hydrogen column density from the
Australia Telescope Compact Array (ATCA) 21 cm measurements by
Stanimirovi\'c et al.\ (1999). The fit was satisfactory in this
scenario, but one can see that the SMC column density necessary to
account for the X-ray absorption is ca.\ two times higher than the
ATCA measurement. In principle, this is not an irreconcilable
difference. The discrepancy can be reasonably explained by the fact
that the ATCA measurements are effectively averaged over a spatial
resolution element of 98 arcsec ($\sim$30 pc at the SMC), while the
X-ray absorption is probing a specific line of sight. It is well known
that the distribution of hydrogen in the SMC is very inhomogeneous
(Staveley-Smith et al.\ 1997; Stanimirovi\'c et al.\ 1999), and the
difference between the ATCA and X-ray fit values for the absorbing
column is not dramatically large. Further uncertainly comes from the
fact that it is possible that the assumed SMC metallicity may not
apply to this specific line of sight, which may affect the estimate of
the X-ray absorbing column. Thus it is in principle possible that the
absorption is in fact caused only by the material in the SMC.

In the second scenario, we fixed the SMC absorbing column at the ATCA
value and attributed the additional absorption to the material at
quasar redshift, assumed to have solar abundances. This scenario gave
a marginally better fit than the first case. The required quasar
intrinsic absorption, $N_{\rm H,Q} = 3.5\times10^{22}$ cm$^2$, is in
the moderate-to-high range, but well within the range seen among AGN
(see, e.g., Malizia et al.\ 1997; Risaliti, Elvis, \& Nicastro 2002).

Both scenarios give virtually identical intrinsic spectral properties
for the quasar. The spectrum appears to be rather typical (e.g.\
Mushotzky, Done, \& Pounds 1993; Fiore et al.\ 1998; Reeves \& Turner
2000). Figure~\ref{fig:0057xspe} shows the X-ray spectrum and the
model fits to the second scenario; the difference between the fits in
both scenarios is too small to be visible on the plot.

We also considered a third scenario, in which both the SMC and
redshifted absorption column densities were allowed to vary, but the
resulting improvement in the fit did not justify the additional degree
of freedom. We need to add that it is, of course, possible that
neither scenario is entirely correct and that there are additional
absorbers on the line of sight to the quasar, possibly a strong
associated absorber or an intervening damped Ly-$\alpha$ system (see,
e.g., Bechtold et al.\ 2001). This possibility cannot be explored with
the available data.

\begin{figure}[ht]
\epsscale{1.0}
\plotone{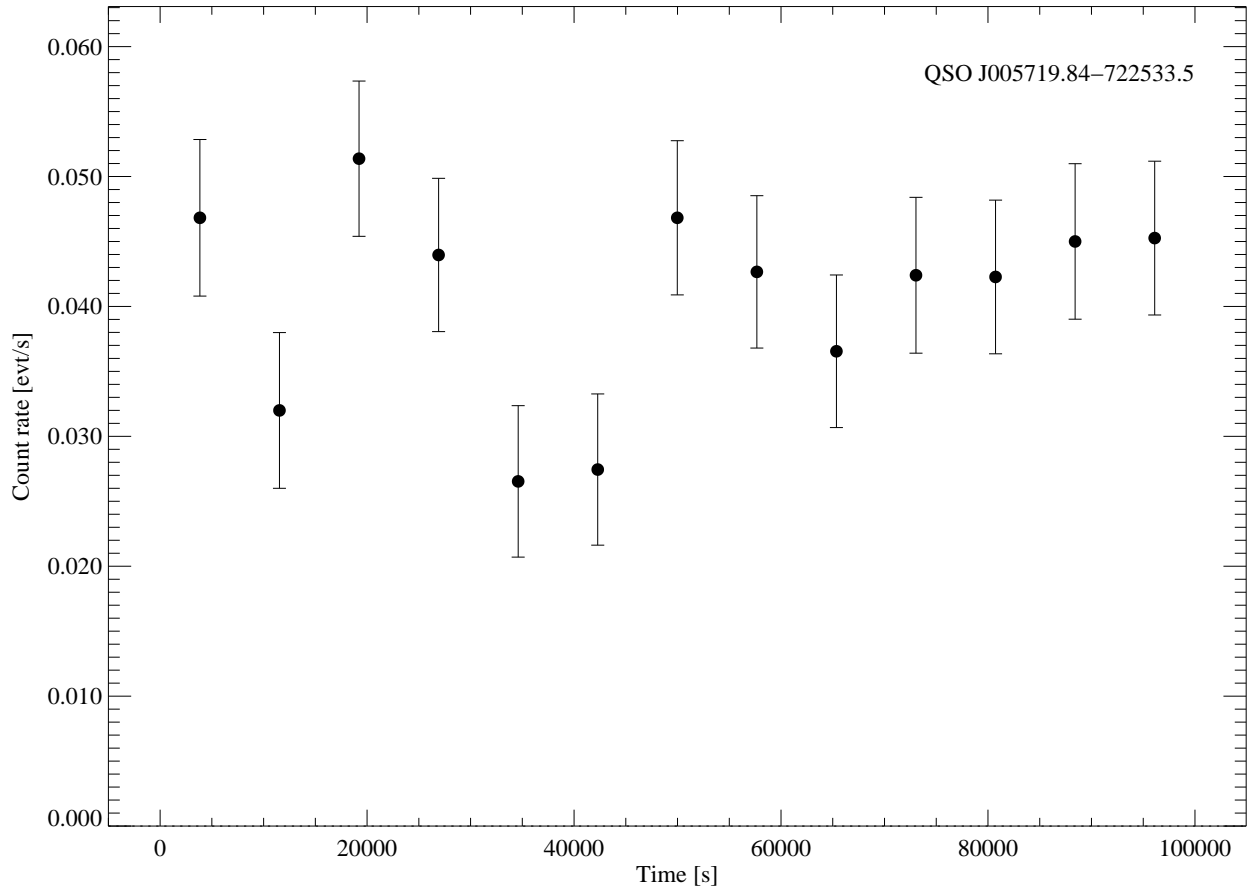}
\caption{X-ray light curve of QSO J005719.84--722533.5. Time tag of 0
seconds corresponds to the beginning of the {\em Chandra\/} observation,
UT 2001 May 15 01:54:13.\label{fig:0057xlc}}
\end{figure}

The X-ray spectral fits were very good, but we note that there are
some features seen in the fit residuals, suggesting that the spectrum
may be more complex. In particular, we note that there is a small hump
near $E=5.6$~keV, which is $\sim$6.4~keV in the quasar rest frame,
i.e.\ just where one would expect to see the iron K$\alpha$ line. We
note, however, that adding a Gaussian line to the modeled spectrum did
not improve the fit.

It is well established that quasars are often variable in X-rays on
short timescales (e.g.\ Mushotzky, Done, \& Pounds 1993; Fabian et
al.\ 2002). QSO J005719.84--722533.5 is no exception; it is variable
at very high confidence level. We show its X-ray light curve in
Figure~\ref{fig:0057xlc} (we note that in this plot we included events
with $E<0.3$~keV, since low energy calibration uncertainties are not
relevant to this issue). During the ca.\ 28 hours in which the quasar
was continuously observed, its ACIS-S3 count rate varied by a factor
of two; we note that the background light curve shows no signs of
variability during the observation. The characteristic timescale for
X-ray variability is $\sim$2.5~hours, again well within the typical
range for quasars.

\subsection{Quasars in the vicinity of SNR \SNRE.}

As mentioned above, four of the quasars lie in the vicinity of the
{\em Chandra\/} calibration source, SNR \SNRE. This object has been
repeatedly observed every few months during the course of the mission,
giving a unique opportunity to study variability of the quasars in the
timescales of the order of months (see, e.g., Nandra et al.\ 1997;
Bauer et al.\ 2002; Maoz et al.\ 2002; Uttley et al.\ 2003).

\begin{figure}
\epsscale{0.76}
\plotone{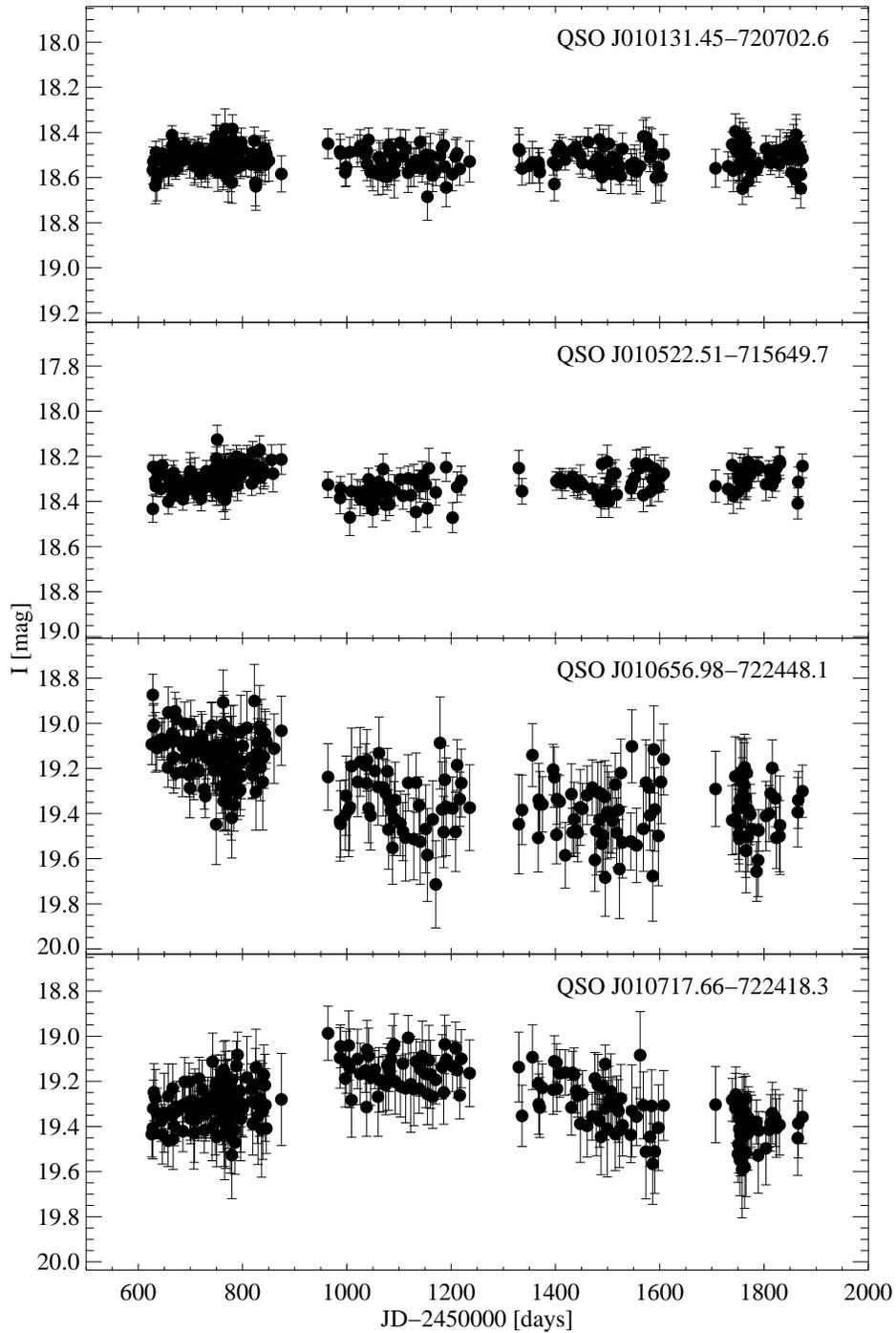}
\caption{OGLE light curves of four quasars in the vicinity of {\em
Chandra\/} calibration source, SNR \SNRE. JD 2,450,000 corresponds to
UT 1995 October 9. Data provided by A. Udalski.\label{fig:e0102ogle}}
\end{figure}

Neither of the quasars was classified by OGLE as variable and were
therefore not considered in the variability-based search for quasar
candidates of Eyer (2002). However, post factum analysis of the OGLE
light curves (Figure~\ref{fig:e0102ogle}; the data were kindly
provided to us by A.~Udalski) qualitatively suggests that at least
some of them are in fact variable.

Since in various {\em Chandra\/} observations \SNRE\ was placed on
various ACIS chips at various off-axis angles, it was a matter of
coincidence whether any of the quasars happened to be in the field of
view in any single observation, and which ACIS CCD chip it happened to
be observed with. As a result, the X-ray data for any of the objects
is a mix of data taken at various off-axis angles at various readout
chips and even various nodes within one chip. Also, the sources were
usually observed far off-axis, where the Chandra's PSF is large, which
forced us to use large source regions, and we typically have a
substantial number of background events in the source regions. In
addition to that, the selection of background regions is severely
constrained, in both spatial and temporal domains. The background
counts had to be taken from regions with comparable responses to the
source regions, excluding regions containing all known sources, and
they had to be observation-specific. All those effects result in
rather large uncertainties for derived source parameters, especially
when temporal properties of the sources are analyzed.

For all quasars, we followed a similar procedure: we extracted data
from source and background regions in all observations in which the
quasar was in the field of view. We performed the spectral analysis
using SHERPA, simultaneously fitting a single model to all
datasets. Again, we excluded photons with energies below 0.3~keV, and
we fitted the spectrum assuming the intrinsic quasar spectrum to be a
power law, with fixed Galactic absorption and additional absorption
from the SMC. In one case we also investigated the possibility that
the absorbing material could be associated with the quasar.

We note that the errors quoted in Table~\ref{tab:x} are formal
uncertainties as determined from the spectral fits. However, in some
cases there may be hard to quantify, systematic effects. In
particular, the estimates of the absorbing column density may be
affected by the fact that many of the {\em Chandra\/} observations of
our quasars were done with the frontside-illuminated ACIS CCD chips,
which have low quantum efficiency in the energy range where absorption
is primarily felt, below $\sim$1~keV.

We analyzed the long-timescale variability of the quasars by grouping
the observations that occurred within few days from one another and
performing individual spectral fits on those groups. The group
selection was straightforward, since the calibration observations were
always done in batches that were separated by several months. Since
the number of events in such groups was often small, we used a
simplified procedure for spectral fits. We froze the spectral slope
and absorbing column and allowed only the spectrum normalization
amplitude to vary.

Notes about the individual objects follow.

\subsubsection{QSO J010131.45--720702.6}

This is a $I=18.5$ quasar at $\zem=0.64$. Even though it has the
noisiest optical spectrum of all objects presented here, its quasar
nature seems certain. The spectrum contains two broad features, near
4580 and 7970~\AA, identified as \ion{Mg}{2} and H$\beta$.

There are no previously known X-ray sources in the vicinity of this
object.

This object was in the {\em Chandra}/ACIS field of view in 27
observations. It is, however, faint in X-rays. Even though the
combined exposure time for this object was 211.5 ksec, it yielded only
212 net photons with energies above 0.3 keV. This enabled only very
limited X-ray spectral analysis.

We fixed the X-ray spectral slope photon power law index to the
canonical value of 1.7 and the SMC absorption to the Stanimirovi\'c et
al.\ (1999) value, $4.09\times10^{21}$ cm$^{-2}$, and we only allowed
the spectrum normalization to vary. We list the results in
Table~\ref{tab:x}.

We would like to note that the observed spectrum contains relatively
few soft photons, and that we get much better fit if we set the
absorption column to values much higher than the value quoted
above. It is, of course, possible that the intrinsic X-ray spectrum is
harder than the assumed $\Gamma=1.7$, which would allow for lower
absorbing column. However, even for the photon index as low as 1.2 we
get the best fit for the column density $\sim$50\%\ higher than the
ATCA value. This may indicate that there are other sources of
absorption on the line of sight to the quasar.

\begin{figure}
\epsscale{0.76}
\plotone{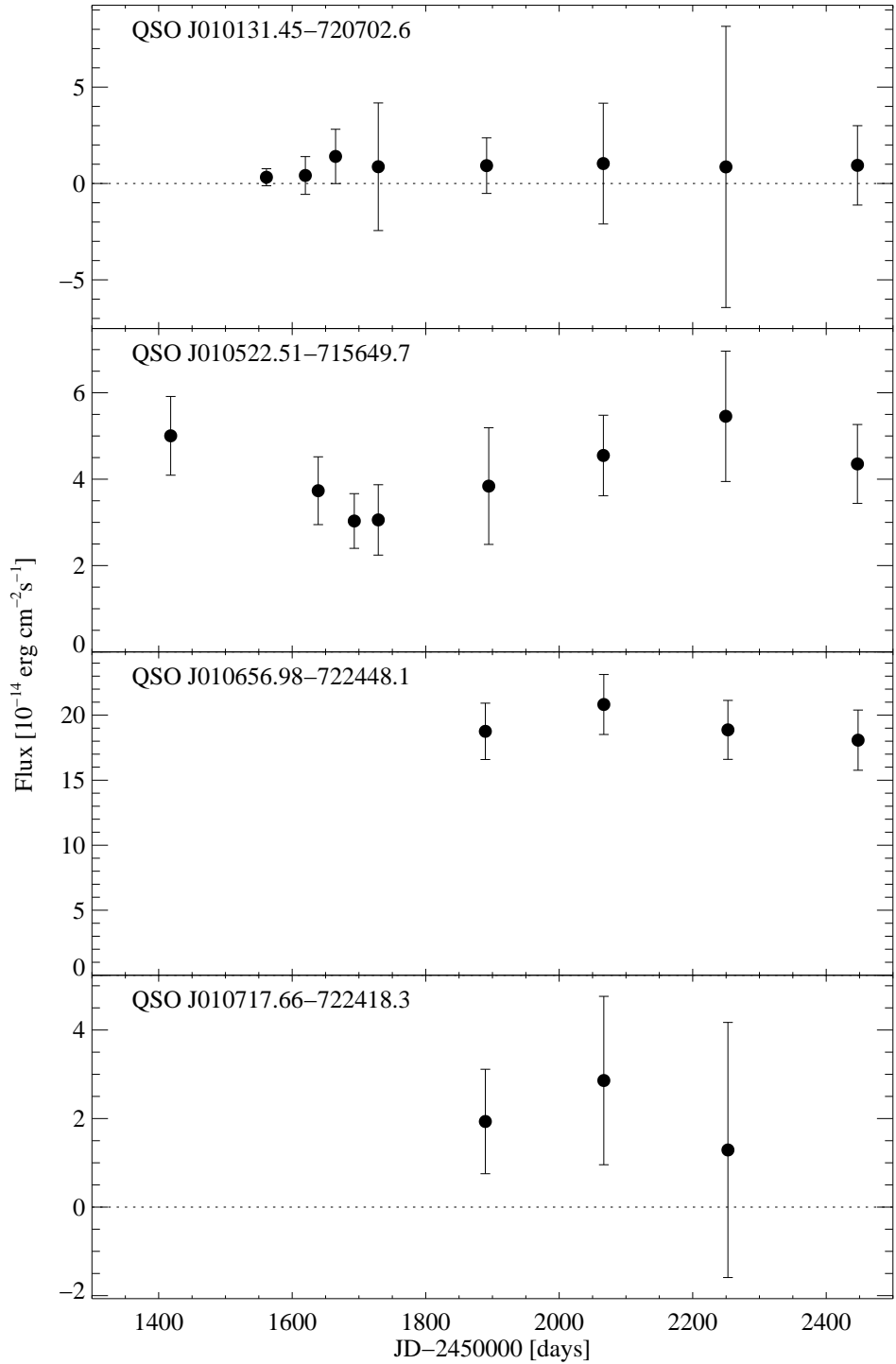}
\caption{X-ray light curves of four quasars in the vicinity of {\em
Chandra\/} calibration source, SNR \SNRE.  JD 2,450,000 corresponds to
UT 1995 October 9. Note that the timescale shown on this plot is much
longer than the timescale from
Fig.~\ref{fig:0057xlc}.\label{fig:e0102xlc}}
\end{figure}

Not surprisingly, individual data points on the X-ray light curve have
large uncertainties (see Figure~\ref{fig:e0102xlc}), driven by low
signal-to-noise ratio in individual observations. The quasar is
clearly too faint to draw any far reaching conclusions, but we can
qualitatively say that it shows no indication of variability.

\subsubsection{QSO J010522.51--715649.7}

This is a $\zem=1.64$ quasar with $I=18.3$. No X-ray sources were
previously known in its vicinity. It was observed by {\em Chandra\/}
36 times for the total of 278~ks, yielding 1166 source photons with
$E\geq0.3$ keV.

In this particular case the X-ray dataset is not well suited for
determination of the absorbing column, since vast majority (32 out of
36) of {\em Chandra\/} observations of this quasar were done with the
frontside-illuminated ACIS CCD chips. We thus fixed the SMC absorbing
column at the value from Stanimirovi\'c et al.\ (1999) and only
allowed the power law index and normalization to vary. The results of
the X-ray spectral analysis are shown in Table~\ref{tab:x}. While
still within the typical range, this quasar shows a somewhat steep
X-ray spectrum.

QSO J010522.51--715649.7 has the best long baseline time coverage of
all the quasars in our sample.  While its X-ray light curve
(Fig.~\ref{fig:e0102xlc}) hints at variability in the 1-1.5 year
timescales, with the amplitude variation of 30-40\%, we note that this
result is not highly significant.

\subsubsection{QSO J010656.98--722448.1}

This $\zem=0.51$ quasar is the second brightest in X-rays of the
quasars presented in this paper. It is however faint in the
optical/near IR range ($I=19.2$). The quasar entered the {\em
Chandra}/ACIS field of view in just ten observations, in four epochs,
with the combined exposure time of $\sim$75 ksec. The observations
yielded 1160 source photons with $E\geq0.3$ keV.

QSO J010656.98--722448.1 was observed in X-rays with {\em ROSAT\/}
(Tab.~\ref{tab:x}), but was not identified as a quasar.

The quasar has very typical X-ray properties. The SMC absorbing column
in the fit is in good agreement with the ATCA measurement; the small
discrepancy can be reasonably explained by the clumpiness of hydrogen
content in the SMC. We also attempted a fit with the SMC column
density fixed at the ATCA value allowing for intrinsic absorption at
the quasar. Since the result from the SMC-only scenario is close to
the ATCA value to begin with, the resulting quasar absorption is ---
not surprisingly --- only weakly constrained.

Since the object has been observed in X-rays in only four epochs,
there is limited variability information available, even though the
quasar is relatively bright in X-rays and individual measurements are
reliable. We note that the light curve (Fig.~\ref{fig:e0102xlc}) is
statistically consistent with a constant flux.

\subsubsection{QSO J010717.66--722418.3}

This is the quasar with highest redshift, $\zem=1.81$, but it is faint
in both the optical range ($I=19.3$) and in X-rays. It was observed by
{\em Chandra\/} only seven times, in only three epochs, with the
combined exposure of $\sim$52 ksec. The total yield was 142 source
photons with energies higher than 0.3 keV, allowing for only very
limited spectral analysis. We fixed the spectral slope at 1.7 and the
SMC absorption at the ATCA value, $3.55\times10^{21}$ cm$^{-2}$, and
only allowed the spectrum normalization to vary. The result is listed
in Table~\ref{tab:x}. For completeness, we show the X-ray light curve
of the quasar in Fig.~\ref{fig:e0102xlc}. The quasar is clearly too
faint for any meaningful variability analysis; we only note that the
data are consistent with no variability.

The position of the quasar agrees very well with the position of a
weak, unclassified source observed in X-rays with {\em ROSAT\/}
(Haberl et al.\ 2000).

\section{Discussion}

We present five X-ray-selected quasars behind the dense parts of the
Small Magellanic Cloud, increasing the number of known such quasars by
$\sim$40\%. Their positions make them excellent reference points for
the proper motion studies. The efficiency of our quasar detection
method in the SMC is comparable to the efficiency seen earlier in the
LMC (Dobrzycki et al.\ 2002).

Only one of the five quasars was classified by OGLE as a variable
object. This is a clear confirmation of the fact that X-ray-based
quasar searches are complimentary to variability-based searches (Eyer
2002; Dobrzycki et al.\ 2003; Geha et al.\ 2003).

The quasars presented here show typical X-ray properties. It was
feasible to analyze X-ray absorption in only two of the available
lines of sight (of the other objects, two were too faint in X-rays and
one had little usable data below 1~keV). In one case, QSO
J005719.84--722533.5, there appears to be some evidence that the
absorbing column in the SMC may not be enough to explain the observed
dip in the spectrum at low energies and that there is additional
source of absorption, possibly associated with the quasar. In the
second case, QSO J010656.98--722448.1, SMC absorbing column alone
seems to satisfactory explain the observed absorption.

One of the five quasars --- QSO J005719.84--722533.5 --- is
particularly interesting. At $I=17.5$ it is sufficiently bright in the
optical/near IR range to allow future spectroscopy with the Cosmic
Origin Spectrograph aboard the Hubble Space Telescope. It is also
relatively bright in X-rays and it could be the quasar that would
provide direct determination of the distance to the SMC. Ideally, such
a quasar would show a quick, in the timescale of days, rise in X-ray
flux. That would enable measurement of the growth of the angular size
of the X-ray halo scattered on the dust in the SMC and a geometrical
determination of the distance to the Cloud. Such rapid rises in X-ray
brightness are not unheard of (see, e.g., Maoz et al.\ 2002; Uttley et
al.\ 2003). Currently, we only have a direct evidence of
short-timescale variability in X-rays for this quasar. However, there
are indications that at least in some cases long-timescale variations
in X-rays and in the optical range do correlate (Peterson et al.\
2000; Uttley et al.\ 2003; but see also Maoz et al.\ 2002). This
quasar is optically variable with the characteristic timescale of
several days (Figure~\ref{fig:0057ogle}), and it is continuously
monitored by the OGLE project. It is thus viable to set up X-ray
monitoring based on the behavior of the optical brightness of the
quasar.


\acknowledgments

We thank B. Draine, L. Eyer, B. Paczy\'nski, and A. Siemiginowska for
helpful discussions, A. Udalski for providing us with unpublished
data, and the referee for very helpful comments. This research has
made use of the Chandra Data Archive (CDA), which is the part of the
Chandra X-Ray Observatory Science Center (CXC), operated for NASA by
the Smithsonian Astrophysical Observatory, of the NASA/IPAC
Extragalactic Database (NED), operated by the Jet Propulsion
Laboratory, California Institute of Technology, under contract with
NASA, and of the SIMBAD database, operated at CDS, Strasbourg,
France. AD acknowledges support from NASA Contract No.\ NAS8-39073
(CXC). LMM was supported by the Hubble Fellowship grant HF-01153.01-A
from the Space Telescope Science Institute, which is operated by the
Association of Universities for Research in Astronomy, Inc., under
NASA contract NAS5-26555.



\begin{deluxetable}{ccccccccccccc}
\rotate
\tabletypesize{\tiny}
\tablenum{1}
\tablewidth{0pt}
\tablecaption{Quasar properties.\label{tab:x}}
\tablehead{%
\colhead{OGLE ID\tablenotemark{a}} &
\colhead{Offset\tablenotemark{b}} &
\colhead{$I$\tablenotemark{c}} &
\colhead{Redshift} &
\colhead{Exp\tablenotemark{d}} &
\colhead{$N_X$\tablenotemark{e}} &
\colhead{$N_{\rm H,G}$\tablenotemark{f}} &
\colhead{$N_{\rm H,ATCA}$\tablenotemark{g}} &
\colhead{$N_{\rm H,X}$\tablenotemark{h}} &
\colhead{$N_{\rm H,Q}$\tablenotemark{i}} &
\colhead{$\Gamma$\tablenotemark{j}} &
\colhead{Norm\tablenotemark{k}} &
\colhead{Flux\tablenotemark{l}}}
\startdata
J005719.84--722533.5\tablenotemark{m} & 0.7 &
17.5 & 0.15 &  98.7 & 2816 & 0.675 & 7.38 &
14.6$\pm$1.5 & \nodata & 2.08$\pm$0.11 & 11.3$\pm$1.1    & 24.7 \\
                                      & &
     &      &       &      &       &      &
ATCA FIX     & 35.3$\pm$7.3 & 2.02$\pm$0.10 & 10.62$\pm$0.87    & 25.4 \\
 & & & & & & & & & & & & \\
J010131.45--720702.6 & &
18.5 & 0.64 & 211.5 &  212 & 0.707 & 4.09 &
ATCA FIX      & \nodata & 1.7 FIX       & 0.214$\pm$0.025 & 0.86 \\
 & & & & & & & & & & & & \\
J010522.51--715649.7 & 0.7 & 
18.3 & 1.64 & 278.2 & 1166 & 0.628 & 6.19 &
ATCA FIX      & \nodata & 1.63$\pm$0.04 & 0.959$\pm$0.041  & 4.28 \\
 & & & & & & & & & & & & \\
J010656.98--722448.1\tablenotemark{n} & 0.4 &
19.2 & 0.51 &  75.3 & 1160 & 0.540 & 3.51 &
5.19$\pm$1.21 & \nodata & 1.45$\pm$0.10 & 3.19$\pm$0.30   & 19.2 \\
                                      & &
     &      &       &      &       &      &
ATCA FIX & 1.18$\pm$0.77 & 1.45$\pm$0.09 & 3.18$\pm$0.27   & 19.2 \\
 & & & & & & & & & & & & \\
J010717.66--722418.3\tablenotemark{o} & 0.6 &
19.3 & 1.82 &  52.5 & 142  & 0.515 & 3.55 &
ATCA FIX      & \nodata & 1.7 FIX       & 0.555$\pm$0.065 & 2.24 \\
\enddata
\tablenotetext{a}{Contains J2000.0 equatorial coordinates.}
\tablenotetext{b}{Offset between {\em Chandra\/} and OGLE positions, in
arcsec.}
\tablenotetext{c}{$I$ magnitude, from \.Zebru\'n et al.\ 2001
(J005719.84--722533.5) and Udalski et al.\ 1998 (remaining four
objects).}
\tablenotetext{d}{{\em Chandra\/} exposure time in ks. For the last four
objects it is the combined exposure for all observations in which
it was in the field of view.}
\tablenotetext{e}{Net source events with $E\geq0.3$ keV after
background subtraction.}
\tablenotetext{f}{Galactic absorbing column in $10^{21}$ cm$^{-2}$, from
http://cxc.harvard.edu/toolkit/colden.jsp}
\tablenotetext{g}{SMC absorbing column in $10^{21}$ cm$^{-2}$, from
ATCA observations by Stanimirovi\'c et al.\ 1999. Values are
$\pm$0.01.}
\tablenotetext{h}{SMC absorbing column in $10^{21}$ cm$^{-2}$, from
spectral fit. ``ATCA FIX'' indicates the value fixed at
the Stanimirovi\'c et al.\ 1999 value.}
\tablenotetext{i}{Intrinsic quasar absorbing column in $10^{21}$
cm$^{-2}$, from spectral fit.}
\tablenotetext{j}{Photon spectral index, from spectral fit. ``FIX''
indicates that the value was fixed.}
\tablenotetext{k}{Power law normalization at 1 keV,
in $10^{-5}$ photons cm$^{-2}$s$^{-1}$keV$^{-1}$, from spectral fit.}
\tablenotetext{l}{Unabsorbed flux in the 2-10 keV range, in $10^{-14}$
erg cm$^{-2}$s$^{-1}$, from spectral fit.}
\tablenotetext{m}{Object No.\ 120 in Kahabka et al.\ 1999 (classified
there as a ``weak hard X-ray binary''), object No.\ 234 in Haberl et
al.\ 2000, and object No.\ 18 in Yokogawa et al.\ 2000.}
\tablenotetext{n}{Object No.\ 200 in Kahabka et al.\ 1999 (classified
there as a ``weak hard X-ray binary''), object No.\ 230 in Haberl et
al.\ 2000 (classified there as ``hard''), and object No.\ 116 in
Sasaki et al.\ 2000.}
\tablenotetext{o}{Object No.\ 227 in Haberl et al.\ 2000.}
\end{deluxetable}



\clearpage

\begin{references}

\reference{} Bauer, F.~E., et al.\ 2002, preprint (astro-ph/0212389)

\reference{} Bechtold, J., Siemiginowska, A., Aldcroft, T.~L., Elvis,
M., \& Dobrzycki, A.\ 2001, \apj, 562, 133

\reference{} Dobrzycki, A., Groot, P.~J., Macri, L.~M., \& Stanek,
K.~Z.\ 2002, \apjl, 569, L15

\reference{} Dobrzycki, A., Macri, L.~M., Stanek, K.~Z., \& Groot,
P.~J.\ 2003, \aj, 125, 1330

\reference{} Eyer, L.\ 2002, AcA, 52, 241

\reference{} Fabian, A.~C.~et al.\ 2002, \mnras, 335, L1

\reference{} Fiore, F., Elvis, M., Giommi, P., \& Padovani, P.\ 1998,
\apj, 492, 79

\reference{} Garmire, G., Feigelson, E.~D., Broos, P., Hillenbrand,
L.~A., Pravdo, S.~H., Townsley, L., \& Tsuboi, Y.\ 2000, \aj, 120,
1426

\reference{} Geha, M., et al.\ (MACHO collaboration) 2003, \aj, 125, 1

\reference{} Haberl, F., Filipovi\'c, M.~D., Pietsch, W., \& Kahabka,
P.\ 2000, \aaps, 142, 41

\reference{} Hamuy, M., Walker, A.~R., Suntzeff, N.~B., Gigoux, P.,
Heathcote, S.~R., \& Phillips, M.~M. 1992, \pasp, 104, 533

\reference{} Kahabka, P., Pietsch, W., Filipovi\'c , M.~D., \& Haberl,
F.\ 1999, \aaps, 136, 81

\reference{} Klose, S.\ 1994, \apjl, 423, L23

\reference{} Malizia, A., Bassani, L., Stephen, J.~B., Malaguti, G.,
\& Palumbo, G.~G.~C.\ 1997, \apjs, 113, 311

\reference{} Maoz, D., Markowitz, A., Edelson, R., \& Nandra, K.\
2002, \aj, 124, 1988

\reference{} Mushotzky, R.~F., Done, C., \& Pounds, K.~A.\ 1993,
\araa, 31, 717

\reference{} Nandra, K., George, I.~M., Mushotzky, R.~F., Turner,
T.~J., \& Yaqoob, T.\ 1997, \apj, 476, 70

\reference{} Paczy\'nski, B.\ 1991, AcA, 41, 257

\reference{} Peterson, B.~M., et al.\ 2000, \apj, 542, 161

\reference{} Predehl, P., Burwitz, V., Paerels, F., \& Tr{\" u}mper,
J.\ 2000, \aap, 357, L25

\reference{} Reeves, J.~N., \& Turner, M.~J.~L. 2000, \mnras, 316, 234

\reference{} Risaliti, G., Elvis, M., \& Nicastro, F.\ 2002, \apj,
571, 234

\reference{} Russell, S.~C., \& Dopita, M.~A.\ 1990, \apjs, 74, 93

\reference{} Russell, S.~C., \& Dopita, M.~A.\ 1992, \apj, 384, 508

\reference{} Sasaki, M., Haberl, F., \& Pietsch, W.\ 2000, \aaps, 147,
75

\reference{} Stanimirovi\'c, S., Staveley-Smith, L., Dickey, J.~M.,
Sault, R.~J., \& Snowden, S.~L.\ 1999, \mnras, 302, 417

\reference{} Staveley-Smith, L., Sault, R.~J., Hatzidimitriou, D.,
Kesteven, M.~J., \& McConnell, D.\ 1997, \mnras, 289, 225

\reference{} Tr{\"u}mper, J.~\& Sch{\"o}nfelder, V.\ 1973, \aap, 25,
445

\reference{} Udalski, A., Szyma\'nski, M., Kubiak, M., Pietrzy\'nski,
G., Wo\'zniak, P., \& \.Zebru\'n, K.\ (OGLE collaboration) 1998, AcA,
48, 147

\reference{} Uttley, P., Edelson, R., McHardy, I.~M., Peterson, B.~M.,
\& Markowitz, A.\ 2003, \apjl, 584, L53

\reference{} Yokogawa, J., Imanishi, K., Tsujimoto, M., Nishiuchi, M.,
Koyama, K., Nagase, F., \& Corbet, R.~H.~D.\ 2000, \apjs, 128, 491

\reference{} \.Zebru\'n, K., et al.\ (OGLE collaboration) 2001, AcA,
51, 317

\end{references}
\end{document}